\documentclass[doublecol]{epl2} 

\title{{Phase Field Crystals as a Coarse-Graining\\ in Time of Molecular Dynamics}}

\author{P. F. Tupper\inst{1} \and Martin Grant\inst{2}}
\shortauthor{Tupper and Grant}

\institute{                    
  \inst{1} Department of Mathematics and Statistics, McGill University, Canada.\\
  \inst{2} Department of Physics, McGill University, Canada.
}

\abstract{ Phase field crystals (PFC) are a tool for simulating
materials at the atomic level.     They combine the small length-scale
resolution of molecular dynamics (MD) with the ability to simulate
dynamics on mesoscopic time scales. 
We show how PFC can be 
{interpreted as the result of}
applying coarse-graining 
 in time to the microscopic density field of molecular dynamics simulations.
{We take the form of} the free energy for the phase field from the classical
density functional theory of inhomogeneous liquids {and then choose coefficients to match the structure factor of the time coarse-grained microscopic density field.}
As an example, we show how to  {construct a} PFC free energy 
for Weber and Stillinger's two-dimensional
square crystal potential which models a system of proteins
suspended in a membrane.
}

\begin{document}

\maketitle

\section{Introduction}
Phase field crystals (PFC) were introduced in \cite{mg1}
as a tool for performing long-time simulation of materials at the
microscopic scale.  The approach is an example of the \emph{phase
  field} approach to non-equilibrium modeling and simulation.  The
system of interest is represented by a field corresponding to the
atomic density of the material.  Its dynamics are simulated
dissipatively according to free energy minimization with additional noise. 
  The advantage of PFC over other phase field models is its ability to
  naturally resolve microscopic structure: the PFC field corresponding to 
 a solid in equilibrium has a periodic structure matching that of the
 crystal stucture of the material.
The advantage of PFC over molecular dynamics (MD) is that fast degrees
of freedom are not present in the model.   This feature allows efficient
simulation over mesocopic time scales. 
The PFC approach can model elastic and plastic deformation of crystals
 as well as the liquid-solid transition \cite{mg2}  and diffusion of
 defects \cite{berry}.

Previous work \cite{mg2}
 demonstrated how phase field crystals could be
constructed that matched the first peak in the structure factor of a
two-dimensional hexagonal crystal 
material and demonstrated their use for the simulation of a 
variety of non-equilibrium phenomenon.
Here we will accomplish two things: (i) show that
phase-field crystals can be interpreted as the result of
applying coarse-graining  in time to  microscopic 
molecular dynamics
(ii) show how to generalize phase-field crystals in order to model
 different materials and match more than the first peak in the
 structure factor. 
The system we will use as an example for this is Weber and
Stillinger's \cite{stillweb}  
two-dimensional  square-crystal forming system.  This system was
originally proposed as a model of various systems where square
two-dimensional crystals occur, including proteins suspended in a
membrane and planar levels in colloidal suspensions.
We will show that we can obtain a PFC free energy that matches
 arbitrarily many peaks of the stucture factor of the time-coarsened
 microscopic density for this potential.

\section{Coarse-Graining Molecular Dynamics}
We consider a system of $N$ particles in the plane each with mass $m$
and with positions
$r_i$ and momenta $p_i$, $i= 1, \ldots,N$.
The energy of the system is given by 
\begin{equation}
H_N = \sum_{i=1}^N \frac{p_i^2}{2m} + V(r_1,\ldots,r_N).
\end{equation}
The $V$ used by Weber and Stillinger
 \cite{stillweb} contains both two-body and three-body
terms:
\begin{eqnarray*}
V(r_1, \ldots, r_N) & = & \sum_{i \neq j} v^{(2)}(|r_i-r_j|) +
  \lambda \sum_{i,j,k} v^{(3)}(r_i,r_j,r_k).
\end{eqnarray*}
  The pair potential is given by
\begin{eqnarray*}
v^{(2)}(r) & =  & \left\{
\begin{array}{ll}
A(r^{-12} -r^{-5} )\exp[(r-a)^{-1}], & 0 <r <a \\
0, &                               r\geq a,
\end{array} \right.
\end{eqnarray*}
where $A=6.767441, a=2.46491832$.  The three-body term is given by 
\begin{eqnarray*}
 v^{(3)}(r_i,r_j,r_k) &  = &    h(r_{ij},r_{jk},\theta_i) +
  h(r_{ji},r_{jk},\theta_j)  \\
  & + & h(r_{ki},r_{kj},\theta_k),
\end{eqnarray*}
where $r_{ij}= |r_i-r_j|$ and $\theta_i$ is the angle between
  $r_j-r_i$ and $r_k-r_i$.  The function $h$ is
\begin{eqnarray*}
h(r,s,\theta)  &=&
\exp[(r-a_3)^{-1} + (s-a_3)^{-1}] \sin^2(2 \theta), 
\end{eqnarray*}
when $r,s < a_3$, and is zero otherwise.
Weber and Stillinger choose $a_3=1.8$ and $\lambda=5$.
  Without the three-body terms the potential is qualitatively
the same as the Lennard-Jones potential, and the system has a
hexagonal crystal phase like that modeled by the PFC in \cite{mg2,berry}. 
The addition of the three-body term makes hexagonal structure
unfavourable and the crystal phase has square symmetry \cite{stillweb}.
The dynamics of the system were simulated using molecular dynamics.

\begin{figure}
\includegraphics[width=1.7in]{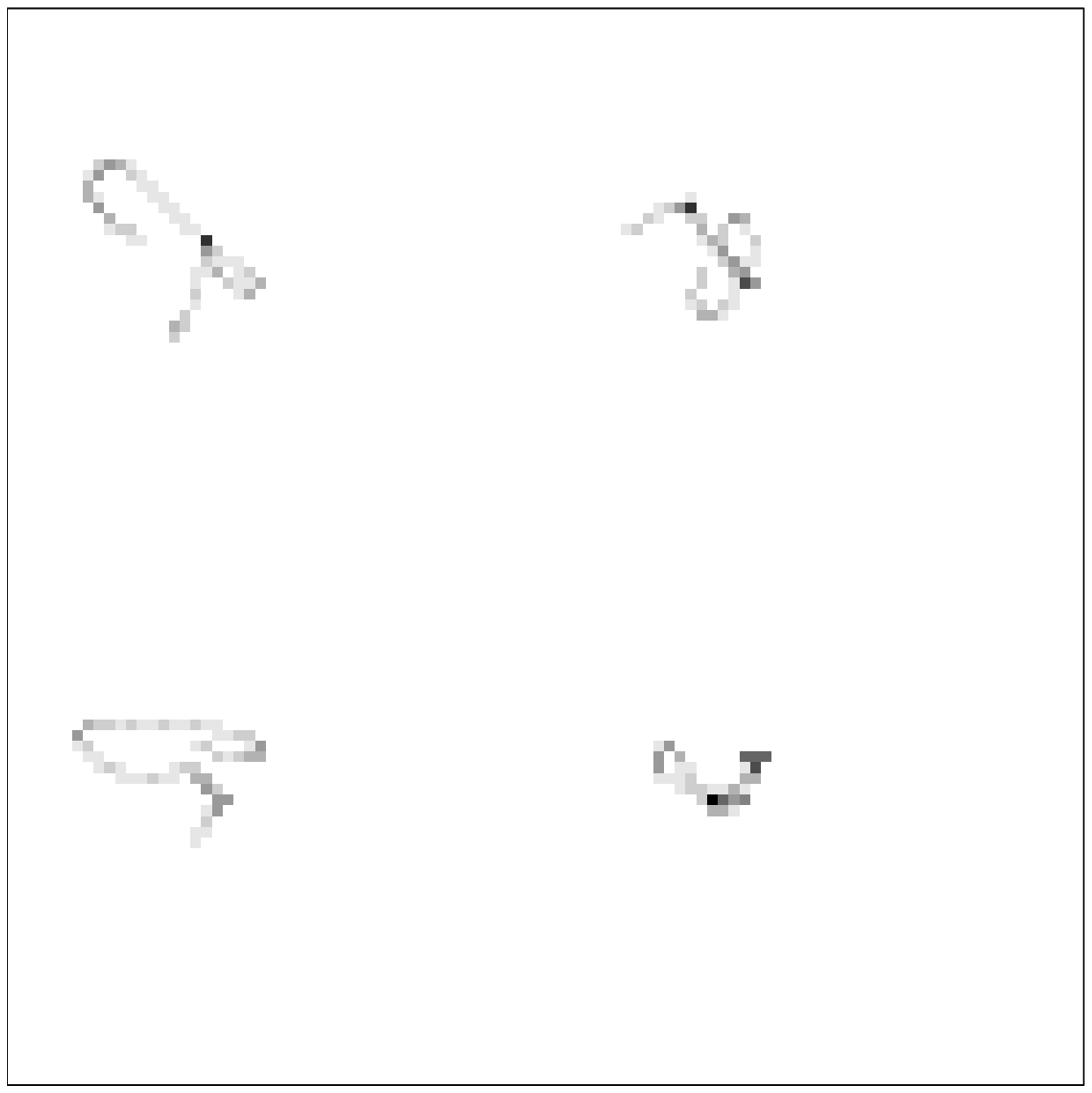}
\includegraphics[width=1.7in]{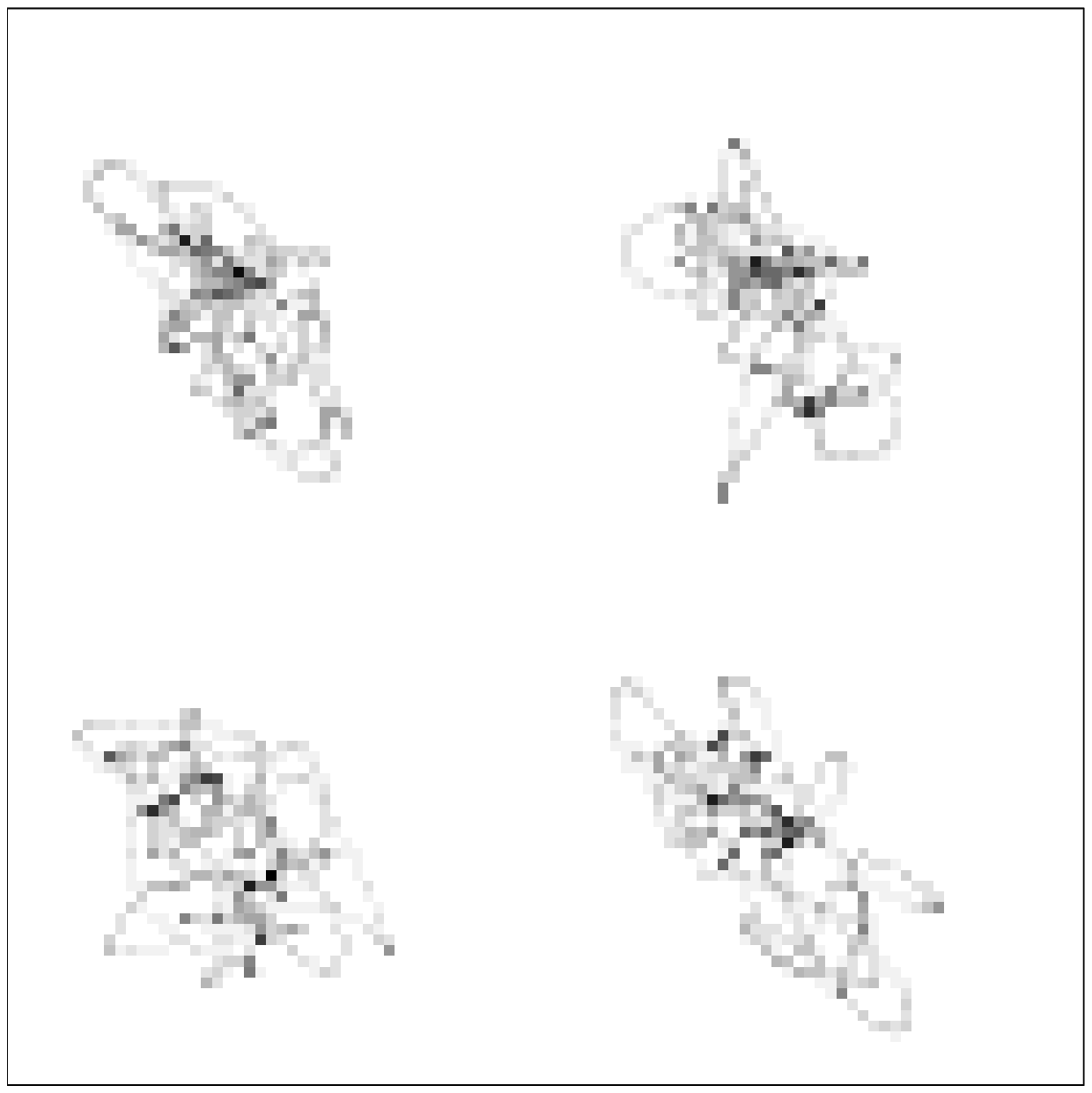}\\
\includegraphics[width=1.7in]{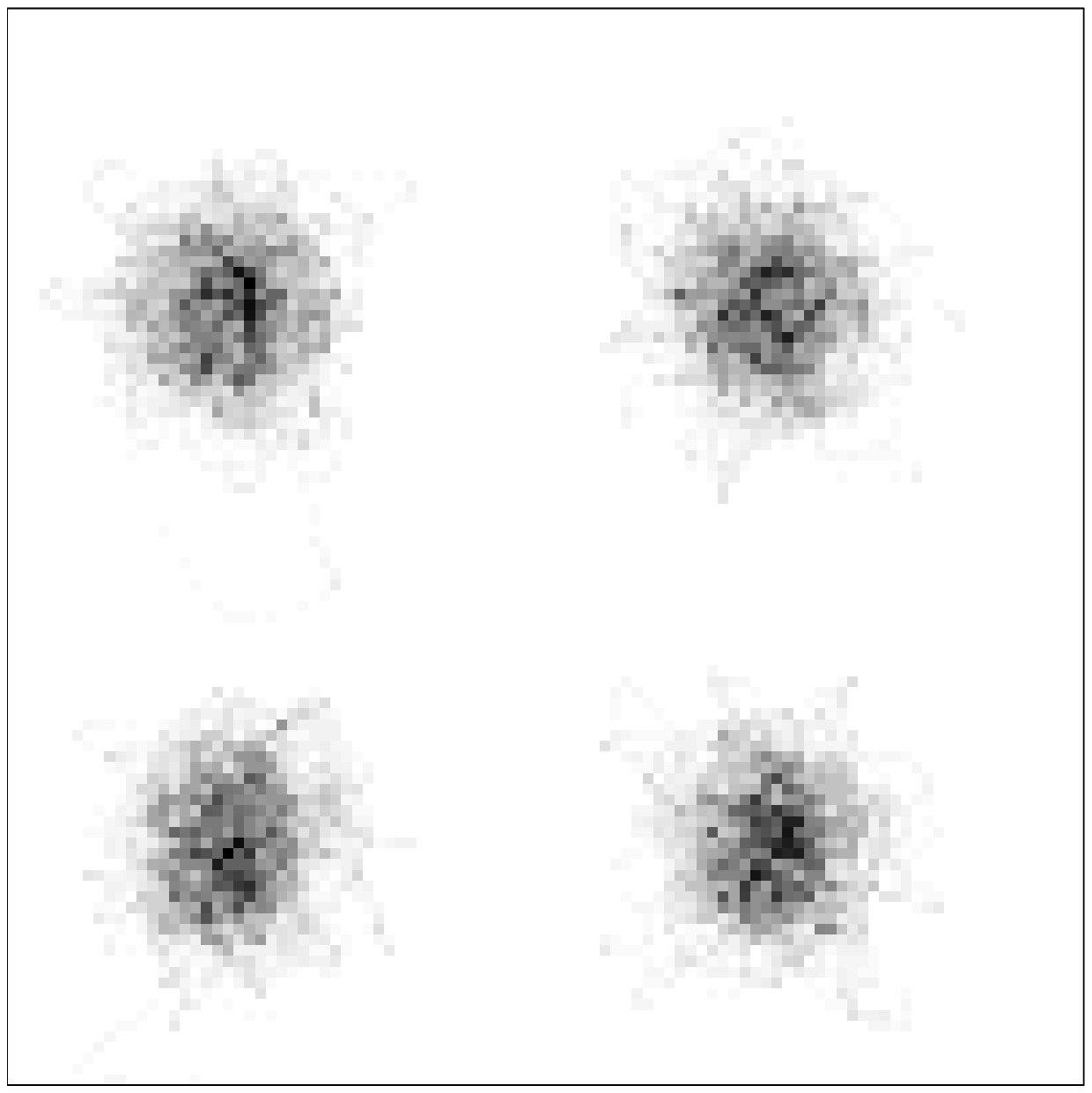}
\includegraphics[width=1.7in]{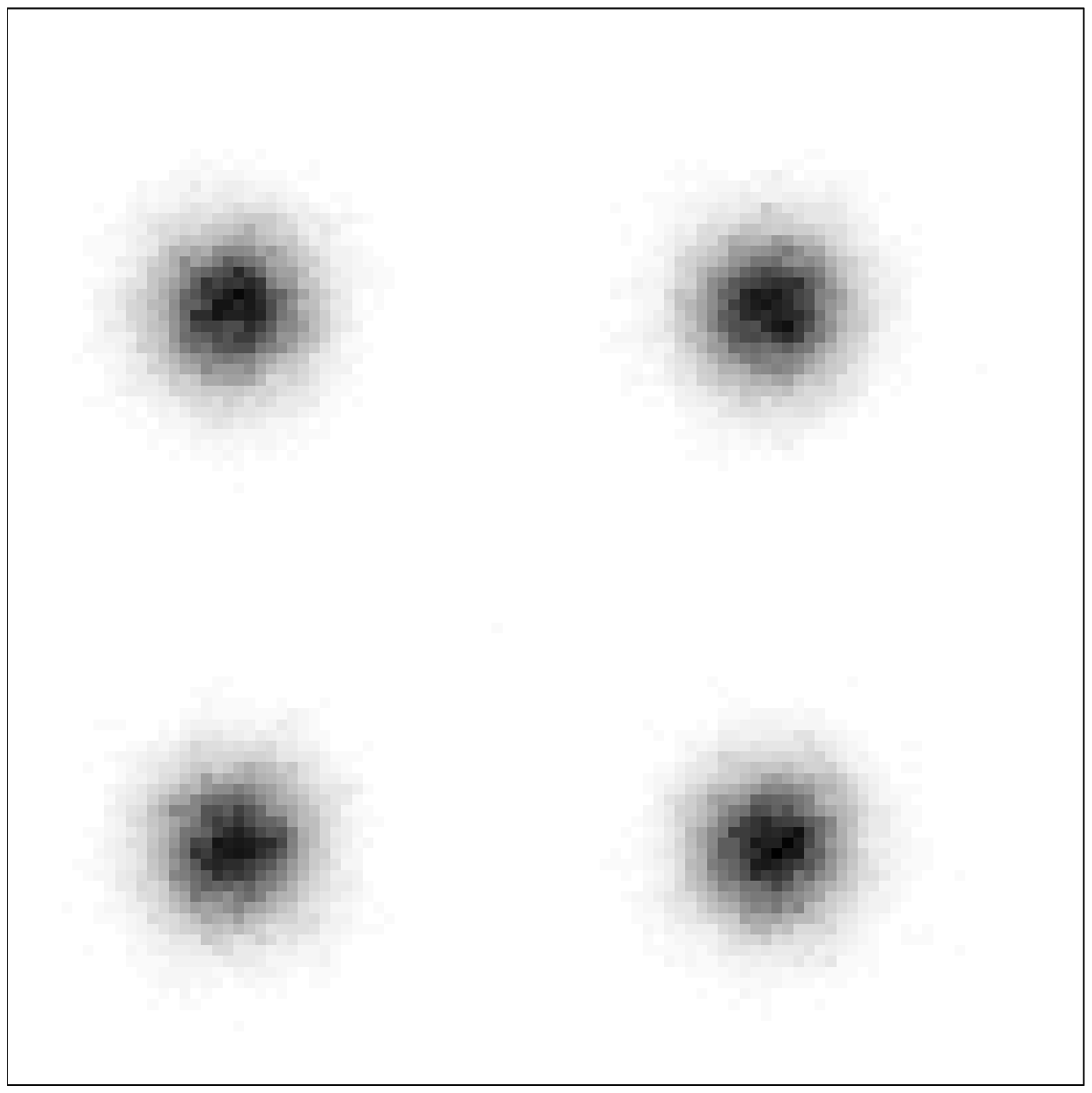}
\caption{Grey-scale plot of  time-averaged density of
  portion of Weber--Stillinger MD 
  system for reduced temperature 0.6, reduced density 0.77277.  The
  time intervals for averaging are 1, 10, 100, 1000 time units.}
\label{fig:renorm}
\end{figure}

The microscopic density field is given by
\(
\rho_m(r,t) = \sum_{i=1}^{N} \delta( r- r_i(t)),
\)
where $r_i(t)$ is the position of the $i$th particle at time $t$.
We consider the density coarse-grained in time:
\begin{equation} \label{eqn:rhodef}
\rho(r)  = \frac{1}{\tau} \int_0^\tau dt\  \rho_m(r)
\end{equation}
If the original system is in the liquid phase, for large $\tau$ the field $\rho$
will be equal to a constant plus small, spatially homogenous
fluctuations.  If the original system is in the solid phase, $\rho$
will have a steady periodic form for large $\tau$ corresponding to the
structure of the crystal.
Fig.~\ref{fig:renorm} shows density plots of $\rho$ of a system in
solid phase for four 
 different coarse-graining times $\tau$.  As $\tau$ is increased
 fluctuations about equilibrium become smaller and occur over a
 smaller time scale.

\section{The Phase Field Crystal}
Our goal is to obtain a phase field model for the dynamics of the coarse-grained
density field $\rho$ given by eq.~(\ref{eqn:rhodef}) for large but finite $\tau$.
{This goal can be broken into two distinct parts.  Firstly, we need to determine a free energy $\mathcal{F}$  functional whose minimum is a good approximation to the equilibrium configuration of $\rho$.  Secondly, we need to specify evolution equations that describe how the system approaches equilibrium.}

\subsection{Free Energy}
The motivation for the particular free energy functional $\mathcal{F}$
we use comes
from classical density functional theory \cite{RY,haymet}.  The goal
of Ramakrishnan 
and Yussouff's  density functional theory  
is to estimate a free energy $\mathcal{F}$ for
which the infinite-time averaged density $\rho= \rho_\infty =
\langle \rho_m \rangle$ is a minimum.
They derive this free energy formally by integrating the partition
function; the calculation can only be completed in the case of the
ideal gas.  To obtain usable approximations  they 
postulate that $\rho$ has a free energy $\mathcal{F}(\rho) =
\mathcal{F}_{id}(\rho) + \mathcal{F}_{ex}(\rho)$ where
$\mathcal{F}_{id}$ is the ideal gas free energy and $\mathcal{F}_{ex}$
is a functional Taylor series expansion in $\rho$ about the uniform liquid
state density \cite{RY}.  Typically, only two terms in the expansion are taken
for most applications \cite{singh}.

Our goal is somewhat different than that of density functional theory; we are
interested in densities averaged over finite times $\tau$, rather
than infinite times.  This
allows us to model non-equilibrium phenomena.
We start with free energies of density functional theory  form and then add
noise to account for our merely finite length of time-coarsening interval $\tau$.
Moreover, rather than deriving coefficients in the free energy from liquid
state equilibrium correlation functions, we choose coefficients that
will be easy to compute, give the correct qualitative behaviour and
have parameters that can then be tuned to match MD.

Our free energy is of the form
\begin{eqnarray*} \label{eqn:freeenergy}
\mathcal{F}(\rho) &=& \int dr T \rho(r) (\log(\rho) - 1) 
   - \int dr \rho(r) \mu \\
  &+& \frac{k_A}{2} \int \! \int   
dr_1 dr_2 C_2(r_1,r_2) \rho(r_1) \rho(r_2) \\
  &+& \frac{k_B}{3} \int \! \! \int \! \! \int dr_1  dr_2 dr_3 C_3(r_1,r_2,r_3)
 \rho(r_1) \rho(r_2) \rho(r_3) 
 \end{eqnarray*}
The last cubic term is necessary to obtain a square crystal structure
and is analogous to the three-body potential term in \cite{stillweb}.
The variable $T$ corresponds to temperature.   

 {According to density functional theory, the kernal $C_2$ should be a multiple of  the direct pair correlation function of the system in the liquid state \cite{chaikin}.  Following this we choose our $C_2$ to be radially symmetic and translationally invariant, giving \( C_2(r_1,r_2) = c(|r_1-r_2|) \) for some function $c$ that decays to zero at infinity.  However, instead of matching $c$ to the correlation function exactly  we instead choose a simple, compact bump function.    Our results are quite insensitive to the exact details of the choice of $c$ we use, the most important quantities being the height and the location of the peak of the Fourier transform of $c$, both of which can be controlled by scaling.  This is predicted by the early papers on density functional theory \cite{RY} and in later analyses tying density function theory to phase field crystals \cite{kenken}.
The function $c(r) =1$ for $|r|<r_0$, $c(r)=0$ otherwise works for our purposes, but in order to minimize grid artefacts we use the following smoother choice:}
\begin{eqnarray*}
 c(r)  & = & \frac{1}{2} \left( \frac{\arctan  20 (r/r_0 - 1) }{2 \pi} -1 \right).
\end{eqnarray*}
For ease of computation, this function is then truncated at radius $2
r_0$ and shifted down to be continuous.  
We fix $r_0=6$ and $k_A=1$.
{Again, the $C_3$ from density functional theory is a three-point correlation function that could in principle be computed from molecular dynamics.  Instead we choose a simple, smooth, compactly supported $C_3$ that gives the qualitatively correct behaviour:}
\begin{eqnarray*}
C_3(r_1,r_2,r_3)&=& \phantom{+}U(r_{12},r_{13},\theta_{123}) 
                + U(r_{21},r_{23},\theta_{213}) \\
      &  \phantom{=}      &   + U(r_{31},r_{32},\theta_{312}) 
\end{eqnarray*}
where $r_{ij}=|r_j-r_i|$ and $\theta_{ijk}$ is the
angle between $r_j-r_i$ and $r_k-r_i$.  
We let 
\begin{eqnarray*}
U(r_{12},r_{23},\theta) &= & H(r_{12}) H(r_{23}) \sin^2(2 \theta)
\end{eqnarray*}
where
\begin{eqnarray*}
H(r) &= & \exp [ (r-r_{max})^{-1} + (r_{min}-r)^{-1} ],
\end{eqnarray*}
for $r_{min} < r < r_{max}$ and $H(r)= 0$ otherwise.
We fix $r_{min}=6$, $r_{max}=10$ and $k_B=0.05$.  {We did not find that the resulting equilibrium configuration was sensitive to the details of $C_3$.  The purpose of the third-order term is only to make a hexagonal pattern unstable and the square pattern more energentically favourable.}

\subsection{Equations for time evolution}
{
Now that we have a form for the free energy $\mathcal{F}$, we need to state equations for the time evolution of the phase field.  In the present work we do not match non-equilibrium properties of the PFC
to those of MD.  Instead we only want plausible dynamics with the correct equilibrium properties.    In this way we can use the dynamics to sample efficiently from the equiblibrium states.
Accordingly, we use a form of equation from the dynamic density functional theory of interacting Brownian particles \cite{archer}:
}
\begin{equation} \label{eq:condynamics}
\frac{\partial \rho(r)}{\partial t}=
\Gamma  \vec{\nabla} \cdot \left( \rho \vec{\nabla} 
\frac{\delta  \mathcal{F}}{\delta \rho(r)} \right)
+  \vec{\nabla} \cdot \left( \sqrt{2 \sigma^2 \Gamma } 
\sqrt{\rho} \, \, \vec{\eta}(r,t) \right) , 
\end{equation}
where $\mathbf{\eta}(r,t)$ is a mean zero noise field with
\begin{eqnarray*}
\langle \eta_i(r,t) \eta_j(r',t') \rangle & = &  \sigma^2 \delta_{ij}  \delta(r-r') \delta(t-t'). 
\end{eqnarray*}
This yields dynamics in which mass is conserved.  We
will also consider
\begin{equation} \label{eq:nondynamics}
\frac{\partial \rho(r)}{\partial t}=
- \Gamma \rho \frac{\delta \mathcal{F}}{\delta \rho(r)} + \Gamma \sigma^2
 + \sqrt{2 \sigma^2 \Gamma} \sqrt{\rho} \, \, \xi(r,t)
\end{equation}
which corresponds to constant chemical potential. 
Here
\begin{eqnarray*}
\langle \xi(r,t) \xi(r',t') \rangle & = & 
\delta(r-r') \delta(t-t'). 
\end{eqnarray*}
 In both cases the noise should be interpreted in the Ito rather than the Stratonovich sense.
 The $\Gamma \sigma^2$ term {in eq.~(\ref{eq:nondynamics})}
does not represent an external force but is a consequence of using multiplicative noise.
It is necessary for the dynamics to satisfy detailed balance, as can be checked with the Fokker-Planck equation.   Both of these dynamics yield an equilibrium distribution for $\rho$
proportional to $\exp (- \mathcal{F}(\rho)/ \sigma^2 )$. 

{
Naturally,  neither of these equations are expected to correctly capture dynamics far from equilibrium and close to equilibrium they will only be accurate in situations where molecular dynamics is well 
matched by Brownian dynamics. }
  The non-conserved dynamics
has the same equilibrium properites as the conserved dynamics
 and can be simulated
numerically more efficiently, so we shall use them for our studies here.
We will show that for suitably chosen $\sigma$ and parameters in  $\mathcal{F}$ the dynamics of eq.~(\ref{eq:nondynamics}) {has an equilibrium distribution that} matches that 
 of the time-averaged microscopic
density $\rho$, at a given temperature, density, and coarse-graining
 time $\tau$.

{The resulting evolution equations when we use non-conserved dynamics  
eq.~(\ref{eq:nondynamics}) with $\mathcal{F}$ given by (\ref{eqn:freeenergy}) are }
\begin{eqnarray*}
\frac{\partial \rho}{\partial t} & = &  -\Gamma \rho
\left[ T \log \rho - \mu 
 + k_A \mathcal{C}\rho + k_B \mathcal{D}(\rho,\rho)\right] \\
& &
 + \Gamma \sigma^2  +
  \sigma \sqrt{2 \Gamma \rho} \, \, \xi,
\end{eqnarray*}
where $\mathcal{C}\rho(r_1) = \int dr_2 c(r_1-r_2) \rho(r_2)$ and
\begin{eqnarray*}
\mathcal{D}(\rho,\rho)(r_1) & = & \int \! \! \int dr_2 dr_3
C_3(r_1,r_2,r_3) \rho(r_2) \rho(r_3). 
\end{eqnarray*}
Here \(\xi\) is noise white in space and time.

\section{Numerical Results}
To numerically simulate the dynamics $\rho$ is discretized on a
rectangular grid. In the units we use, the distance between two
adjacent points on the grid is 1 unit. 
All integrals are evaluated by straightforward summation on the grid. 
The equation are integrated in time using the explicit Euler method,
typically with a time step of length $\Delta t =0.01$.

For particular values of the parameter $\mu$ and without noise
$(\sigma=0)$ the field $\rho$ has two possible equilibria selected
depending on $T$.   For large $T$ the $\log$ term dominates and the
stable equilibrium is a constant $\rho$ corresponding to a liquid
state.  As $T$ is decreased, this state becomes unstable and the
system has a periodic free energy minimum resembling the fourth plot
in fig.~\ref{fig:renorm}.  With the parameters described here the
distance between two adjacent bumps in the periodic phase is
approximately 8.5 units.

\subsection{Nonequilibrium simulation}
\begin{figure}
\includegraphics[width=1.7in]{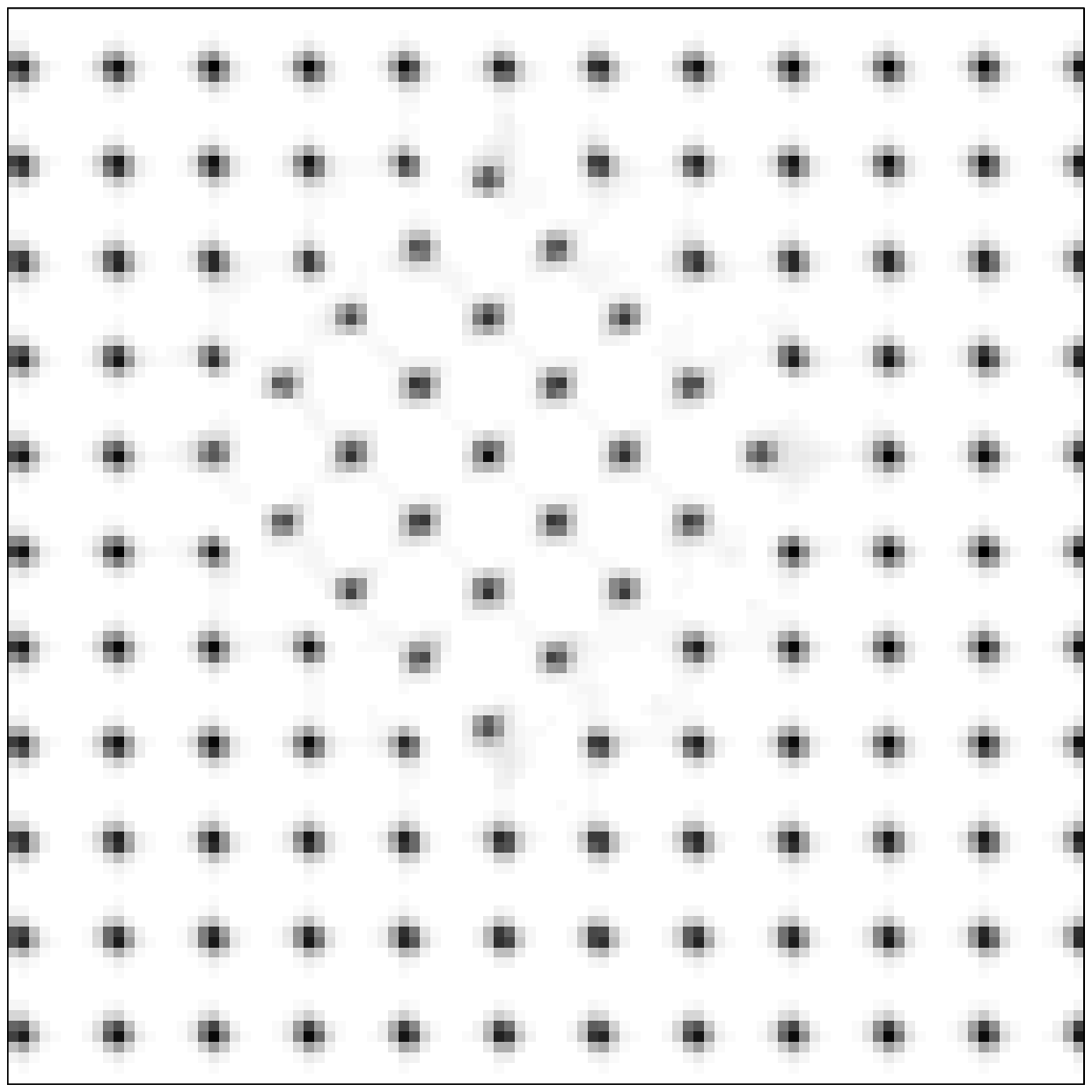}
\includegraphics[width=1.7in]{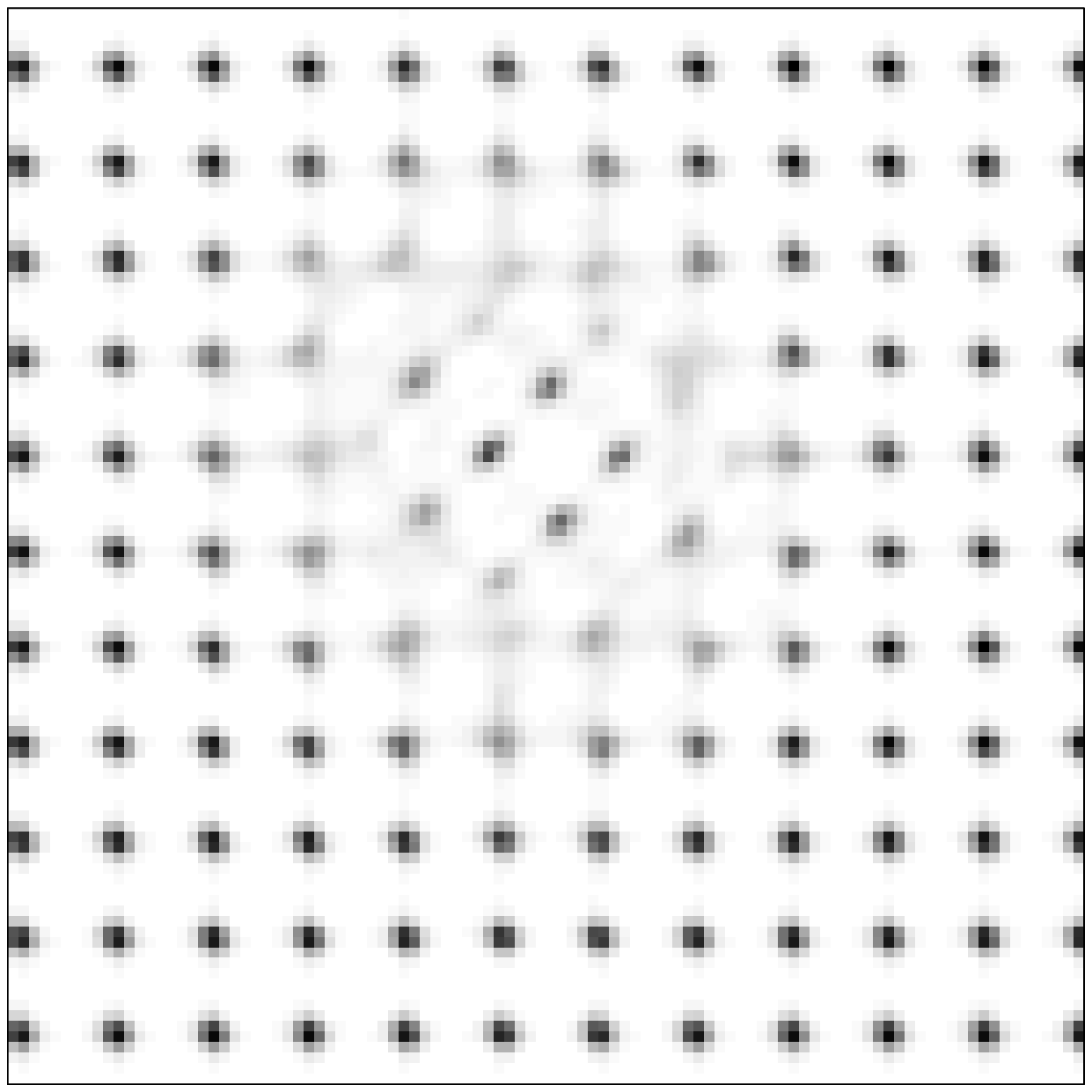} \\
\includegraphics[width=1.7in]{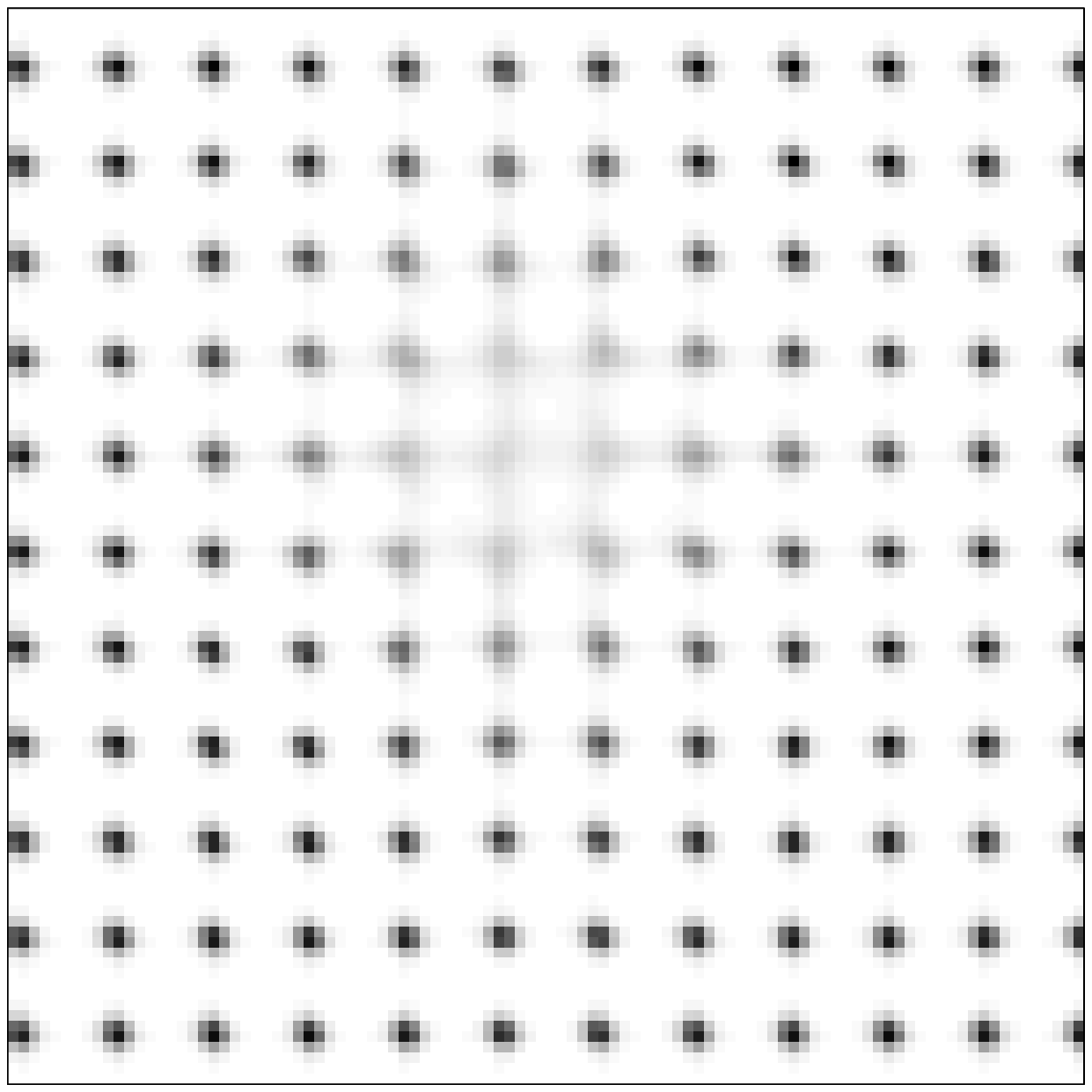}
\includegraphics[width=1.7in]{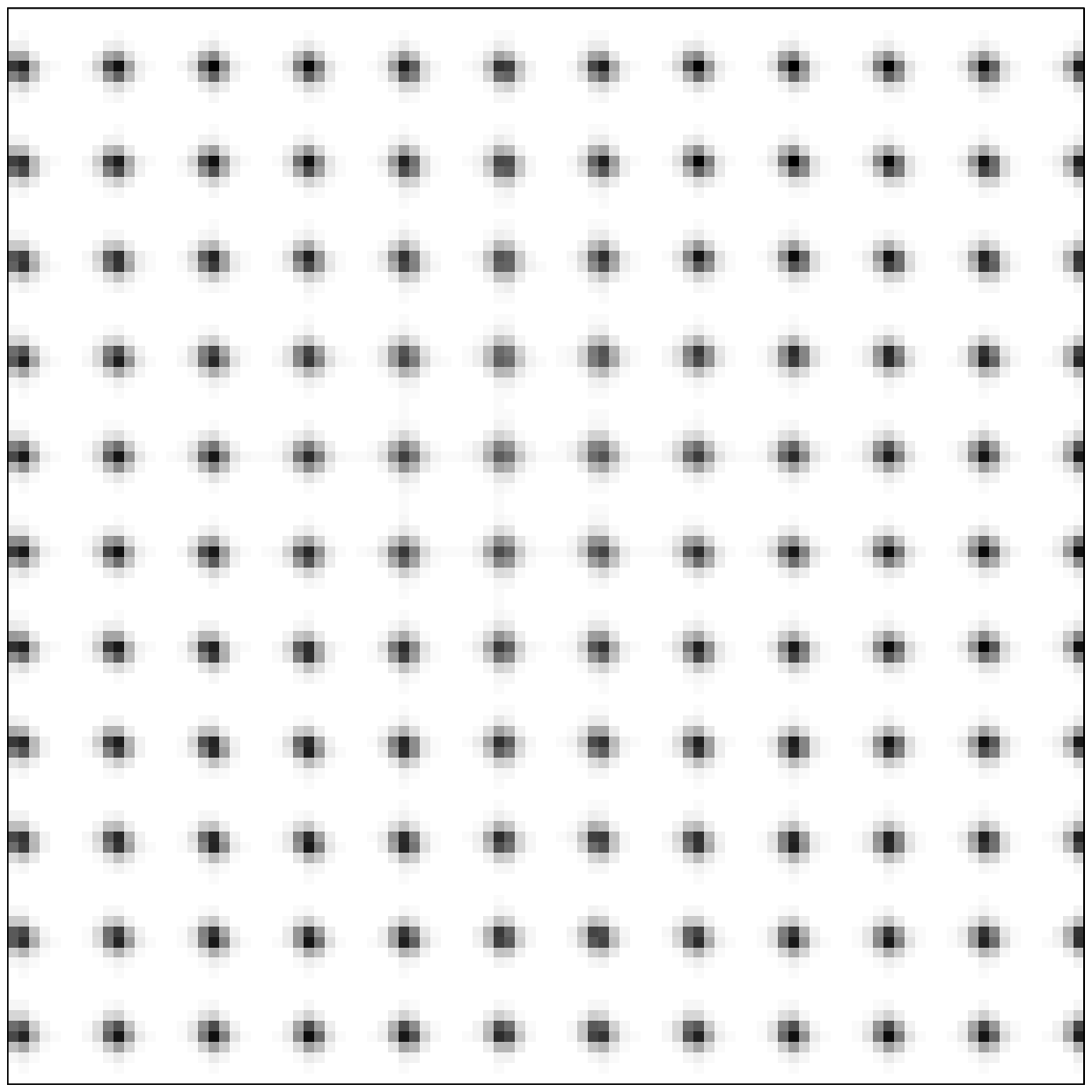}
\caption{ Grey-scale plot of $\rho$ on  a portion of the computational domain
  during simulation with non-conserved dynamics.  Plots shown at
  0, 33, 66, 99 time units.}
\label{fig:grain}
\end{figure}

Fig.~\ref{fig:grain} shows a non-equilibrium simulation of the
 system with non-conserved dynamics. With $\mu=10$, $T=2.2$, $\sigma =0$ the
 system was started in the equilibrium crystal state except for one
 one patch consisting of square of 25 particle bumps which is rotated
 45$^\circ$.   The system was integrated in time until close to the
 uniform solid configuration equilibrium.  In grey-scale we show
 density plots of the system for four representative times.

\subsection{Structure Factor}
We now compare results numerically obtained using this free energy
with those  of time-averaged molecular dynamics.  For a wide
range of temperatures, densities, and time-coarsening times of the MD
system, we can choose 
\(T\), \( \mu \) and \( \sigma \) for the PFC to obtain good
agreement, as we discuss further below.

\begin{figure} 
\includegraphics[width=3.4in]{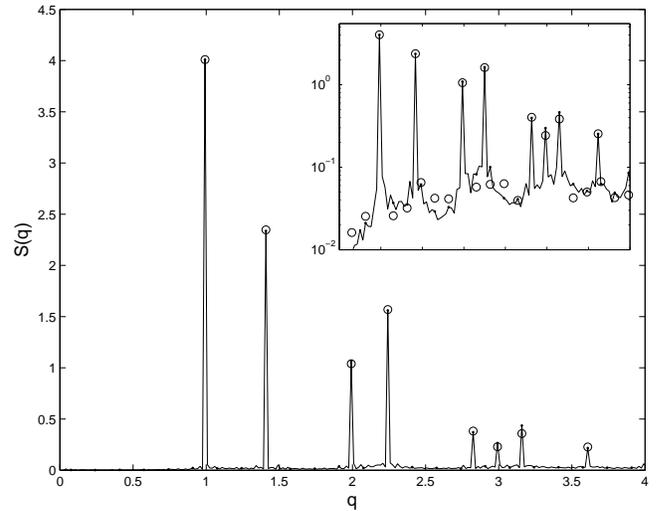}
\caption{ Comparison of structure factor {$S(q)$ versus wavenumber $q$} for MD (line and dots) and PFC
(circles).  For clarity only some of the data points are shown for
  PFC.  The inset shows the same data on a logarithmic scale.}
\label{fig:compmd0p5}
\end{figure}

Fig.~\ref{fig:compmd0p5} shows the structure factor for a
time-averaged MD simulation together with the structure factor for
an equilibrium PFC configuration where parameters have been chosen to
obtain the best fit.  The MD structure factor was taken from a 400
particle system in a perfect square crystal of density 0.77277
(leading to the
minimum energy for the ground state) and
reduced temperature 0.55.  The PFC was obtained from a non-conserved
dynamics run with $\mu=10$, $T=0.8$, $\sigma=0.02$.
{The values for these parameters were chosen as follows.  Since vertical and horizontal scales on the structure factor could be matched trivially be rescaling the dimensions of density and distance, we only concerned ourselves with matching the relative heights of the peaks and their relative spacing.   With $\sigma=0$ (no noise) $\mu$ and $T$ were first chosen in order to obtain a square crystal pattern.  This occurs for a broad range of these parameters and immediately gives the correct spacing of peaks in the structure factor.  No further adjustments of $\mu$ were necessary.  The value of $T$ was then adjusted to match the ratio of the first two peaks of the structure factor to that of the MD results.  As $T$ goes to zero (a perfect crystal)  the relative height of the second peak grows, and as $T$ goes to the melting point the second peak shrinks.   As soon as the second peak was mathced all the remaining peaks agreed to the degree shown in Fig.~\ref{fig:compmd0p5}.  At this point the structure factor of PFC consists purely of peaks and it was necessary to increase $\sigma$ until the background rose to match that of the MD structure factor's background.}

\subsection{Melting Transition}
\begin{figure} 
\includegraphics[width=3.4in]{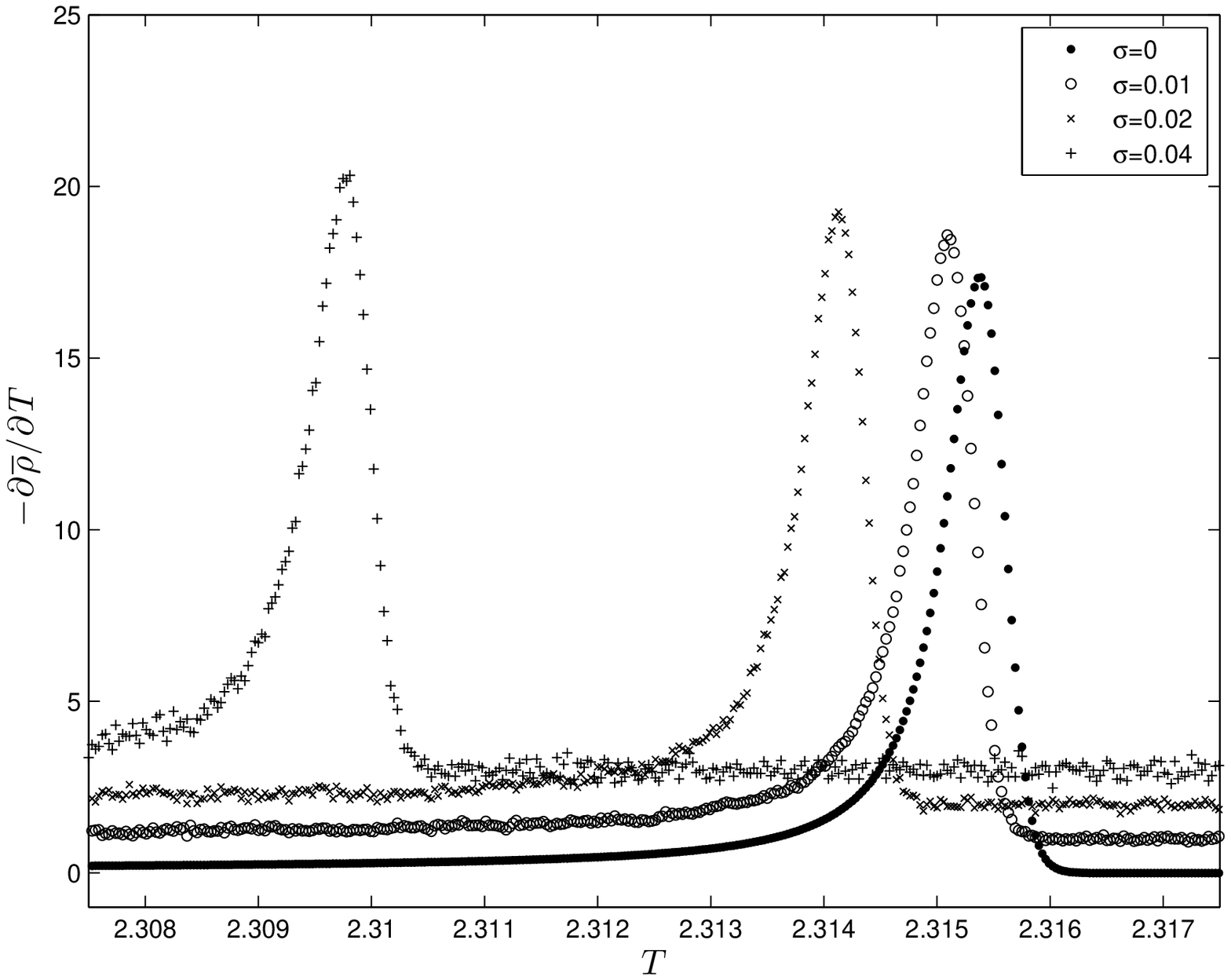}
\caption{Plot of
  $-\partial \bar{\rho} /\partial T$ as a function of $T$ for the PFC
  system, where $\bar{\rho}$ is mean
  density and $T$ is temperature.  
 Results are shown for four
  different values of the noise strength $\sigma$.  Plots are offset
  vertically from zero for clarity.}
\label{fig:pfcphasetrans}
\end{figure}

{An open question in the theory of phase transitions is whether the melting transition form solid to liquid in two-dimensional systems occurs via a single phase transition or if there are two phase transitions with an intermediate hexatic phase \cite{nelson}.  Weber and Stillinger studied this question for their system} using MD,
 and found no
evidence for a two-stage melting 
process \cite{stillweb}. We perform an analogous experiment using our
PFC model of the system.
As in \cite{stillweb}, we examine the liquid-solid transition by
starting the system in a perfect square crystal and increasing the
temperature (in our case, the parameter $T$) slowly.  
We used the non-conserved dynamics with $\mu=30$, and
$\sigma=0,0.01,0.02, 0.04$. 
{A phase transiton should appear in a plot of mean density versus temperature as a discontinuity in the solution or its derivative.   In order to highlight any such points of non-smoothness, we first smooth the mean density as a function of $T$ and then consider its derivative with respect to $T$.}
In fig.~\ref{fig:pfcphasetrans}
we show the negative derivative of mean density $\bar{\rho}$ with respect to $T$ as a function of
$T$.   {We observe only one peak in the derivative for each run of the simulation, suggesting a one-stage melting process.}
 Our results are consistent
with those of Weber and Stillinger.

\acknowledgments
This work was supported by the National Science and Engineering
Research Council and  le Fonds qu\'eb\'ecois de la recherche sur la
nature et les technologies.  The authors thank Ken Elder, Sami
Majaniemi, and Dan Vernon for helpful discussions.


\begin{thebibliography}{0}

\bibitem{mg1}
\Name{K. R. Elder, M. Katakowski, M. Haataja \and M. Grant}
\REVIEW{Phys.\ Rev.\ Lett.}{88}{2002}{245701}.

\bibitem{mg2}
\Name{K. R. Elder \and M. Grant}
\REVIEW{Phys.\ Rev.\ E}{70}{2004}{051605}.

\bibitem{berry}
\Name{J. Berry, M. Grant \and K. R. Elder}
\REVIEW{Phys.\ Rev.\ E}{73}{2006}{031609}.


\bibitem{stillweb}
\Name{T. A. Weber \and F. H. Stillinger}
\REVIEW{Phys.\ Rev.\ E}{48}{1993}{4351}.

\bibitem{RY}
\Name{T. V. Ramakrishnan and M. Yussouff}
\REVIEW{Phys.\ Rev.\ B}{19}{1979}{2775}.

\bibitem{haymet}
\Name{A. D. J. Haymet \and D. W. Oxtoby}
\REVIEW{J.\ Chem.\ Phys.}{74}{1981}{2559}.


\bibitem{singh}
\Name{Y. Singh}
\REVIEW{Phys.\ Rep.}{6}{1991}{351}.

\bibitem{chaikin}
\Name{ P.~M.~Chaikin \and T.~C.~Lubensky}
\Book{Principles of condensed matter physics}
\Publ{Cambridge University Press} 
\Year{1995} \Pages{195}{198}

\bibitem{kenken}
\Name{K. R. Elder, N. Provatas, J. Berry, P. Stefanovic \and M. Grant}
\REVIEW{Phys.\ Rev.\ B}{75}{2007}{064107}


\bibitem{archer}
\Name{A. J. Archer \and M. Rauscher}
\REVIEW{J.\ Phys.\ A: Math.\ Gen.}{37}{2004}{9325}. 


\bibitem{nelson}
\Name{David R. Nelson}
\Book{Defects and Geometry in Condensed Matter Physics}
\Publ{Cambridge University Press}
\Year{2002} \Section{2.4}


\end{thebibliography}
\end{document}